# 2D Be$_3$B$_2$C$_3$: a stable direct-bandgap semiconductor with record-breaking carrier mobility, 8.1× 10$^5$ cm$^2$ V$^{-1}$ s$^{-1}$

Xiao Wang[1*†], Xiaoxin Yang[1,4†], Jiangyu Li[2,3*]

[1]Shenzhen Key Laboratory of Nanobiomechanics, Shenzhen Institute of Advanced Technology, Chinese Academy of Sciences, Shenzhen 518055, China

[2]Department of Materials Science and Engineering, Southern University of Science and Technology, Shenzhen 518055, China

[3]Guangdong Key Provisional Laboratory of Functional Oxide Materials and Devices, Southern University of Science and Technology, Shenzhen 518055, China

[4]Nano Science and Technology Institute, University of Science and Technology of China, Suzhou 215123, China.
†These authors contributed equally: Xiao Wang, Xiaoxin Yang.
*Corresponding author Email: xiao.wang@siat.ac.cn; lijy@sustech.edu.cn.

## ABSTRACT

The Moore's law in the semiconducting industry has faltered as the three-dimensional (3D) Si-based transistors has approached their physical limit with the downscaling. The carrier mobility *μ*, critical to the device's performance, will be degraded when the thickness of Si is scaled into several nanometers. In contrast to the bulk counterpart, two-dimensional (2D) semiconductors can be scaled into atomic-layer thickness without dangling bonds, maintaining its intrinsic carrier mobility and going beyond the limits of Si-based electronics. Hence, the development of novel 2D semiconducting materials with high carrier mobility is the market demand as well as the scientific challenge. Here, we successfully designed 2D Be$_3$B$_2$C$_3$ with planar hypercoordinate motif. It possesses the perfect planar skeleton with both pentacoordinate carbon and hexacoordinate boron moieties, which is the first reported material with such multi-hypercoordinate centers. Density functional theory (DFT) calculations prove that the Be$_3$B$_2$C$_3$ monolayer has excellent structural and thermal stabilities as well as mechanical properties. Further investigations reveal that the Be$_3$B$_2$C$_3$ monolayer has a strong ultrahigh Fermi velocity (2.7 × 10$^5$ m/s), suitable direct bandgap (1.97 eV), and high optical absorption coefficient (10$^5$ cm$^{-1}$). As a result, an unprecedented ultrahigh room-temperature carrier mobility (8.1× 10$^5$ cm$^2$ V$^{-1}$ s$^{-1}$) with strong anisotropy is discovered, making Be$_3$B$_2$C$_3$ monolayer a revolutionary candidate for future electronic and photovoltaic applications.

# INTRODUCTION

Moore's law has been a general criterion to guide the development of semiconducting industry for more than half century, continuously pushing the size scaling of Si-based nanodevices[1,2]. However, in the case of 3D architecture, the carrier mobility, which is vital to the device' performance, decreases exponentially as the size is scaled into several nanometers. 2D materials of which the carriers are confined in their atomically thin layer may go beyond the Si-based electronics[3]. Therefore, the novel design of stable 2D semiconducting materials with excellent carrier mobility is one of the crucial enablers in the "More than Moore" era[4-6]. Up to now, some archetypes or representatives of 2D materials have come to the fore. For instance, graphene[7] is a robust atomically thin 2D carbon sheet with high carrier mobility (~3 × $10^5$ $cm^2$ $V^{-1}$ $s^{-1}$ theoretically)[8]. However, its zero bandgap greatly impedes its application in field-effect devices. Transition metal dichalcogenides (TMDCs)[9] exhibit sizable bandgap (for example, 1.8 eV for $MoS_2$ monolayer), but their application is limited by the low carrier mobility (typically less than 100 $cm^2$ $V^{-1}$ $s^{-1}$ for $MoS_2$ thin flakes at room temperature)[10]. Recently, phosphorene has emerged as a good 2D semiconductor candidate, with both appreciable thickness-dependent bandgap (0.3 to 2.0 eV from bulk to monolayer) and relatively high carrier mobility (~$10^3$ $cm^2$ $V^{-1}$ $s^{-1}$ at room temperature)[4], but it is unstable at ambient environment, greatly hindering its potential application. Therefore, the search for 2D semiconductors with high mobility carrier, appropriate bandgap and good stability remains imperative.

Carbon-based materials, like carbon nanotube[11-15], graphene[16-19], graphdyine[20-22], have displayed high carrier mobility and been explored as the ideal candidates for nanodevices. The planar hypercoordinate carbon motif, derived from the cornerstone of planar tetracoordinate carbon (ptC) molecule[23-28], can provide an ideal and vast carbon-based material-design platform due to the extension of chemical bonding rules[29-34]. Based on the platform, several 2D planar hypercoordinate carbon materials have been discovered. Li *et al*. designed a quasi-planar hexacoordinate carbon (phC)-containing 2D material with a $Be_2C$ stoichiometry[29], and Wang *et al*, on the basis of $Be_9C_2^{4-}$, designed the quasi-planar pentacoordinate carbon-featuring 2D materials $Be_5C_2$[30]. However, the absolutely perfect planar hypercoordinate carbon-based 2D materials with high carrier mobility remains to be discovered.

In this work, we used the "isoelectronics" strategy and designed a novel perfect planar pentacoordinate carbon (ppC)-containing 2D $Be_3B_2C_3$ monolayer by density functional theory (DFT) calculations. Isoelectronic to graphene endows $Be_3B_2C_3$ with excellent structural stability and mechanical properties. Importantly, it exhibits ultrahigh Fermi velocity (2.7 × $10^5$ m/s), suitable bandgap (1.97 eV) and excellent optical absorption coefficient ($10^5$ $cm^{-1}$). And most importantly, it has a theoretically record-breaking ultrahigh root-temperature carrier mobility (8.1 × $10^5$ $cm^2$ $V^{-1}$ $s^{-1}$), guaranteeing $Be_3B_2C_3$ monolayer as a promising semiconductor candidate for future ultrasmall high-performance and low-power electronic nanodevices.

## RESULTS AND DISCUSSION

### *Structure design of Be$_3$B$_2$C$_3$ monolayer and its stability*

Our design of Be$_3$B$_2$C$_3$ monolayer was inspired by isoelectronic consideration (Fig. 1). BeC monolayer is isoelectronic to the boron honeycomb network and formed by substituting boron atoms with beryllium and carbon atoms alternately. However, both the BeC monolayer and boron hexagon are unstable due to the deficiency of two electrons per hexagon compared with graphene (see Supplemental Material for details). In graphene, the valence electron configuration of carbon atom is 2s$^2$2p$^2$, which can form sp$^2$ hybridization and contributes to three local σ-bond and one delocalized π-bond, guaranteeing the hexagon pattern and the stability of graphene. While in boron or BeC honeycomb, the lack of p$_z$ electron in hexagon destabilizes the structure. Therefore, earlier works tried to stabilize the beryllium carbide or boron hexagon by electron donor[29, 35]. For example, Li *et al.* designed the Be$_2$C monolayer by Be doping at the center of each hexagon[29]. However, Be$_2$C monolayer is not a perfect flat but a buckled structure, which might be caused by the electrostatic repulsion between two neighboring beryllium atoms. To circumvent this problem, we introduced boron atom alternating with void in the hexagon to weaken the electrostatic repulsion (Fig. 1), which results in the stoichiometric composition of Be$_3$B$_2$C$_3$ in its primitive cell. In each primitive cell, two boron atoms supply the three Be-C pairs with six electrons, rendering the Be$_3$B$_2$C$_3$ monolayer isoelectronic counterpart of graphene.

As shown in Fig. 2A, our designed Be$_3$B$_2$C$_3$ monolayer is a perfectly planar structure and has a hexagonal skeleton with the optimized lattice constants of a = b = 5.25 Å. The primitive cell of Be$_3$B$_2$C$_3$ monolayer consists of three beryllium atoms, two boron atoms and three carbon atoms. Each carbon atom binds to three beryllium atoms and two boron atoms, forming a perfect ppC moiety that have never been reported before. As a result, a hexacoordinate boron also emerges in the structure, indicating the existence of another multi-hypercoordinate centers. Moreover, there is no energy difference between spin-polarized and spin-unpolarized calculations, suggesting that Be$_3$B$_2$C$_3$ monolayer has a nonmagnetic ground state with no unpaired electron or unsaturated dangling bond. To illuminate the chemical bonds in a more intuitive plot, we calculated the deformation charge density and electron localization function (Fig. 2B, C). There is a remarkable electron donation from beryllium and boron atoms to carbon atoms but a negligible electron transfer between the beryllium and boron atoms, which is consistent with our initial design concept. The transferred electrons are well delocalized around the C-B and C-Be bonds, forming strong covalence that is crucial to the stability of ppC moiety in Be$_3$B$_2$C$_3$ monolayer.

To assess the stability of Be$_3$B$_2$C$_3$ monolayer, we first computed the cohesive energy of Be$_3$B$_2$C$_3$ monolayer, which is defined as $E_{\text{coh}} = \dfrac{\left(hE_{\text{Be}} + kE_{\text{B}} + lE_{\text{C}} - E_{\text{Be}_3\text{B}_2\text{C}_3}\right)}{h + k + l}$, where $E_{\text{Be}}$, $E_{\text{B}}$, $E_{\text{C}}$ and $E_{\text{Be}_3\text{B}_2\text{C}_3}$ represent the energies of a single isolated beryllium, boron and carbon atom and Be$_3$B$_2$C$_3$ monolayer, respectively, and *h*, *k* and *l* represent the number of beryllium, boron and carbon atoms in the primitive cell. The calculated cohesive energy of Be$_3$B$_2$C$_3$ monolayer is 6.27 eV/atom, much higher than that of phosphorene (3.30 eV/atom) and MoS$_2$ (4.98 eV/atom)[29, 36], guaranteeing the strongly connected network in Be$_3$B$_2$C$_3$ monolayer. The kinetic stability of Be$_3$B$_2$C$_3$ monolayer is further confirmed by the phonon dispersion curves (Fig. 2D) without any imaginary mode. We notice that the highest frequency of Be$_3$B$_2$C$_3$ monolayer can reach 1121 cm$^{-1}$, which is higher than those of MoS$_2$ monolayer (473 cm$^{-1}$)[37], silicene (580 cm$^{-1}$)[38] and quasi phC-containing Be$_2$C (1020 cm$^{-1}$)[29], indicating the excellent kinetic stability of Be$_3$B$_2$C$_3$ monolayer. The projected density of states (DOS) reveal that the highest frequency mainly comes from C-Be bonds and C-B bonds (Fig. 2E), demonstrating the great importance of perfect ppC skeleton for the stability of Be$_3$B$_2$C$_3$ monolayer. Besides, the ab-initio molecular dynamic (AIMD) simulations were performed to verify the thermal stability of Be$_3$B$_2$C$_3$ monolayer. We used a 4 × 4 × 1 supercell with time running for 10 ps under 2500 K (Fig. 2F). Interestingly, the structure of Be$_3$B$_2$C$_3$ monolayer is well maintained at such an extreme temperature, demonstrating its high thermal stability and potential high temperature applications.

Mechanical stability is another aspect that reflects the stability of a material. Herein, we investigated the mechanical properties of Be$_3$B$_2$C$_3$ monolayer by examining the elastic constants. The calculated elastic constants are $C_{11} = C_{22} = 157.05$ N/m, $C_{12} = C_{21} = 29.09$ N/m, which meets the Born criteria ($C_{11} > 0$; $C_{11} > |C_{12}|$). We further calculated the direction-dependent Young's modulus, shear modulus, and Poisson's ratio of Be$_3$B$_2$C$_3$ monolayer using a rectangular unit cell (Fig. 2G-I). All the Young's modulus, shear modulus, and Poisson's ratio along the direction are isotropic, and its in-plane stiffness is higher than those of silicene (59 N/m) and MoS$_2$ (121 N/m)[29], suggesting the good mechanic properties of Be$_3$B$_2$C$_3$ monolayer.

*Electronic structure of Be$_3$B$_2$C$_3$ monolayer*

After proving the structure stability, we examined the electronic properties of Be$_3$B$_2$C$_3$ monolayer. We first calculated the band structure by elements weights using Perdew-Burke-Ernzerhof (PBE) functional (see Supplemental Material for details). Be$_3$B$_2$C$_3$ monolayer has direct bandgap of 1.28 eV, having the valence band maximum (VBM) and conduction band minimum (CBM) both located at the Γ point. Considering that the PBE functional may underestimate the bandgap, we also calculated the bandgap of Be$_3$B$_2$C$_3$ monolayer by hybrid Heyd-Scuseria-Ernzerhof (HSE) functional (Fig. 3A), resulting in a 1.97 eV direct bandgap responsive to visible-light.

In order to analyze the orbital contribution to the bands near the Fermi level, we calculated the projected band structures of Be, B and C elements (see Supplemental Material for details), revealing that VBM is mainly contributed by the C-2p$_z$ orbital while the CBM originates mainly from B-2p$_z$ orbital. To demonstrate the projected band structure in a more intuitive plot, we calculated the partial charge densities corresponding to the VBM and CBM (see Supplemental Material for details), which is consistent with the projected band structures. We also investigated the effect of biaxial strain on electronic properties of Be$_3$B$_2$C$_3$ monolayer. The direct bandgap could sustain 10% compressive or tensile strain. Interestingly, the value of the direct bandgap varies almost linearly with the biaxial strain, suggesting that we can achieve precise regulation of bandgap with strain engineering (see Supplemental Material for details).

To deepen the insight into the band edge dispersion in the 3D Brillouin zone, we plotted the 3D band structure of $Be_3B_2C_3$ monolayer, as seen in Fig. 3B-D. Excitingly, the band structure exhibits a strong linear electronic band edge dispersion, which usually indicates a relatively small effective mass. Based on the linear band edge, we calculated the Fermi velocity of $Be_3B_2C_3$ monolayer (Fig. 3E), and the calculated isotropic Fermi velocity $v_F$ can reach $2.7 \times 10^5$ m/s, comparable with graphene ($8.2 \times 10^5$ m/s)[39], which suggests an ultrafast carrier transport in $Be_3B_2C_3$ monolayer.

**Table. 1** The calculated in-plane stiffness ($C$), effective mass ($m^*$), deformation potential ($E_d$) and carrier mobility ($\mu$) of $Be_3B_2C_3$ monolayer at 300 K.

| | Carrier type | Direction | $C$ (N/m) | $m^*$ ($m_0$) | $E_d$ (eV) | $\mu$ (cm$^2$ V$^{-1}$ s$^{-1}$) |
|---|---|---|---|---|---|---|
| $Be_3B_2C_3$ | Hole | $x$ | 157.03 | 0.851 | 0.805 | $7.1 \times 10^3$ |
| | | $y$ | 157.18 | 0.795 | 0.078 | $8.1 \times 10^5$ |
| | Electron | $x$ | 157.03 | 0.237 | 2.352 | $1.0 \times 10^4$ |
| | | $y$ | 157.18 | 0.249 | 1.408 | $2.8 \times 10^3$ |

*Transport and optical properties of $Be_3B_2C_3$ monolayer*

With such interesting structure and electron characteristics, an ultrahigh carrier mobility can be expected. Here, we evaluated the mobility transport within the effective mass approximation and the electron and electron-acoustic phonon scattering mechanism by deformation potential (DP) theory. The carrier mobility ($\mu$) and the affinitive parameters including effective mass ($m^*$), in-plane stiffness ($C$), and deformation potential ($E_d$) of $Be_3B_2C_3$ monolayer are listed in Table 1 (see Supplemental Material for details). It can be seen that the effective masses and in-plane stiffness are isotropic, which is in agreement with the symmetric structure. A small carrier effective mass is obtained at the band edge along the Γ-X and Γ-Y directions, which means a high carrier mobility can be expected. On the basis of the DP theory, we plotted the energy shift of CBM and VBM with respect to the lattice perturbation along $x$ and $y$ directions to obtain the deformation potential $E_d$ (see Supplemental Material for details). The energy shift of CBM and VBM is inconspicuous and hence the calculated $E_d$ will be much smaller. Especially, for the hole along $y$ direction, the $E_d$ can be as small as 0.078, which is about one order of magnitude smaller than other values and dedicates to a larger carrier mobility. We can see that the hole mobility reaches up to $8.1 \times 10^5$ cm$^2$ V$^{-1}$ s$^{-1}$ at room temperature. The mobility at liquid-helium temperature was also calculated to be $6.1 \times 10^7$ cm$^2$ V$^{-1}$ s$^{-1}$. Here we summarized the reported carrier mobility for materials from 0D to 3D (Fig. 4)[4, 16, 20, 40-49], and $Be_3B_2C_3$ monolayer stands out with the highest carrier mobility among all the reported materials. Combining with the high stability and suitable bandgap responsive to visible light, the $Be_3B_2C_3$ monolayer is promising in the high-speed nanodevices such as field effect transistors and memory devices.

The superior electronic properties of $Be_3B_2C_3$ monolayer promise excellent applications in optoelectronic, high-speed electron device and visible-light solar harvesting/utilizing techniques. Hence, the optical performance was investigated further (see Supplemental Material for details). It can be seen that $Be_3B_2C_3$ monolayer has a wide adsorption spectrum from the far-infrared to the ultraviolet region and its adsorption coefficients in visible-light region reach an order of $10^5$ cm$^{-1}$, which is superior to the common semiconductor photovoltaic materials such as Si and MAPbI$_3$[40]. Next, we analyzed the physical reasons behind such superior optical performance using the transition dipole moment (see Supplemental Material for details), and it is found that $Be_3B_2C_3$ monolayer has a strong inter-band optical transition between the CBM and VBM, which are contributed by the boron atoms and carbon atoms, respectively. The robust partial C-B covalent bond in $Be_3B_2C_3$ monolayer builds a bridge to transfer electron easily, which determines the superior optical properties of $Be_3B_2C_3$ monolayer.

## CONCLUSIONS

In summary, by combining the planar hypercoordinate chemistry and 2D architecture, we designed a semiconducting ternary compound $Be_3B_2C_3$ with recording-breaking room-temperature carrier mobility reaching $8.1 \times 10^5$ cm$^2$ V$^{-1}$ s$^{-1}$. Such outbreaking results from collective contribution of strong in-plane stiffness, small effective mass and deformation potential. Besides, this material exhibits high stability, suitable direct bandgap of 1.97 eV and high optical absorption coefficient over wide absorption windows, which presents the remarkable advantage over the commercial electronic and optoelectronic materials. We hope our theoretical prediction will promote the experiment synthesis of this revolutionary material, explore more fantastic properties and make great contributions to the current semiconducting industry. Besides, apart from the simple substance (like graphene, borophene) and common binary compound (like h-BN, MoS$_2$), ternary compounds also provide a great stage for the design of novel 2D materials and semiconducting materials for wide range of potential applications.

## COMPUTATIONAL SECTION

**DFT calculations**. First principles calculations were performed in the framework of the density-functional theory as implemented in the Vienna ab initio simulation package (VASP)[50]. The DFT calculations for the structural relaxation employed the generalized gradient approximation (GGA) formulated by Perdew, Burke, and Ernzerhof (PBE)[51]. Projector augmented wave (PAW) potentials were used in all calculations[52]. A kinetic cutoff energy of 520 eV for the plane wave basis set and a 12 × 12 × 1 k-point mesh was generated[53]. All the ionic relaxations were optimized by the recommended conjugate-gradient algorithm until the maximum atomic force component acting on each atom is less than 0.01 eV/Å. A vacuum space of 15 Å is adopted to minimize the interlayer interaction between adjacent layers. The phonon calculations are performed by the PHONOPY code[54]. The visualization of structure and electron density are implemented in VESTA software[55].

**Mechanical properties.** The angle-dependent Young's modulus, and Poisson's ratio along the in-plane $\theta$ can be obtained from the elastic constants as follows[56]:

$$Y(\theta) = \frac{C_{11}C_{22} - C_{12}^2}{C_{11}\sin^4\theta + C_{22}\cos^4\theta + \left(\frac{C_{11}C_{22} - C_{12}^2}{C_{66}} - 2C_{12}\right)\cos^2\theta\sin^2\theta} \quad (1)$$

$$v(\theta) = -\frac{C_{12}\sin^4\theta + C_{12}\cos^4\theta - \left(C_{11} + C_{22} - \frac{C_{11}C_{22} - C_{12}^2}{C_{66}}\right)\cos^2\theta\sin^2\theta}{C_{11}\sin^4\theta + C_{22}\cos^4\theta + \left(\frac{C_{11}C_{22} - C_{12}^2}{C_{66}} - 2C_{12}\right)\cos^2\theta\sin^2\theta} \quad (2)$$

**Carrier mobility.** The carrier mobility of $Be_3B_2C_3$ can be calculated by means of the deformation potential (DP) theory proposed by Bardeen and Shockley as follows[47]:

$$\mu_{2D} = \frac{e\hbar^3 C_{2D}}{k_B T m^* m_d E_d^2} \quad (3)$$

where $m^*$ and $m_d = \sqrt{m_x^* m_y^*}$ are carrier effective masses along transport direction and average effective mass, respectively. $E_d = \Delta V/(\Delta l/l_0)$ is the deformation potential constant, defined as the shift of band edges induced by strain. And $\Delta V$ represents the energy difference between conduction band minimum and valence band maximum with the lattice applied by proper stretch or compression. $l_0$ and $\Delta l$ are the lattice constants along the $x$ or $y$ direction and their deformation. The in-plane stiffness is derived from $C_{2D} = \left[\partial^2 E/\partial(\Delta l/l_0)^2\right]/S_0$, where $E$ is the total energy, $S_0$ is the area of the optimized supercell. $k_B$ is the Boltzmann constant, $T$ is the temperature. $\hbar = h/2\pi$, where $h$ is the Planck constant.

**Optical properties.** The optical adsorption coefficient $\alpha(\omega)$ can be obtained as follows:

$$\alpha(\omega) = \sqrt{2}\omega\left[\sqrt{\varepsilon_1^2 + \varepsilon_2^2} - \varepsilon_1\right]^{\frac{1}{2}} \quad (4)$$

where the $\varepsilon_1(\omega)$ and $\varepsilon_2(\omega)$ are the real and imaginary parts of the dielectric function, respectively. The imaginary part $\varepsilon_2(\omega)$ of dielectric function can be obtained from the momentum matrix elements between the occupied and unoccupied states as follows:

$$\varepsilon_2(\omega) = \frac{4\pi^2 e^2}{\Omega} \lim_{q \to 0} \frac{1}{q^2} \sum_{c,v,k} 2\omega_k \delta(\varepsilon_{ck} - \varepsilon_{ck} - \omega) \times \langle u_{ck+e_\alpha q} | u_{vk} \rangle \langle u_{ck+e_\beta q} | u_{vk} \rangle^* \quad (5)$$

where $\Omega$ represents the volume, $\alpha$ and $\beta$ are the Cartesian components, $e_\alpha$ and $e_\beta$ are the unit vectors, $v$ and $c$ represent matrix elements of the transition from the valence band state ($u_{vk}$) to the conduction band state ($u_{ck}$), $\varepsilon_{ck}$ and $\varepsilon_{vk}$ denote for the energy of the conduction and valence band, respectively. And the real part $\varepsilon_1(\omega)$ of dielectric function can be derived from the imaginary part using Kramer-Kronig relationship as follows[57]:

$$\varepsilon_1(\omega) = 1 + \frac{2}{\pi} P \int_0^\infty \frac{\varepsilon_2^{\alpha\beta}(\omega')\omega'}{\omega'^2 - \omega^2 + i\eta} d\omega' \quad (6)$$

where $P$ represent the principal value of the integral.

**Molecular dynamic simulations.** The thermal stability of $Be_3B_2C_3$ monolayer was evaluated by means of the ab-initio molecular dynamic (AIMD) simulations. The ground-state structure of $Be_3B_2C_3$ monolayer was annealed at different temperatures. MD simulation in NVT ensemble lasts for 10 ps with a time step of 1.0 fs.

## ACKNOWLEDGEMENTS

This work was supported by the National Natural Science Foundation of China (22003074, 11772207, 12192213), the Youth Innovation Promotion Association CAS, the Guangdong Provincial Key Laboratory Program from the Department of Science and Technology of Guangdong Province (2021B1212040001), Guangdong Provincial Department of Education Innovation Team Program (2021KCXTD012), Shenzhen Science and Technology Program (JCYJ20200109115219157), and Center for Computational Science and Engineering at Southern University of Science and Technology.

**FIGURES AND CAPTIONS**

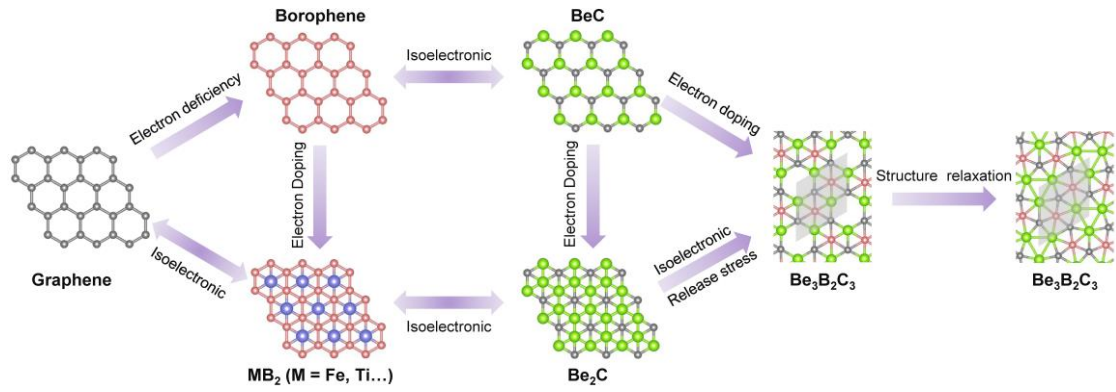

**Figure. 1. Schematic diagram of the design concept of Be$_3$B$_2$C$_3$ monolayer**.

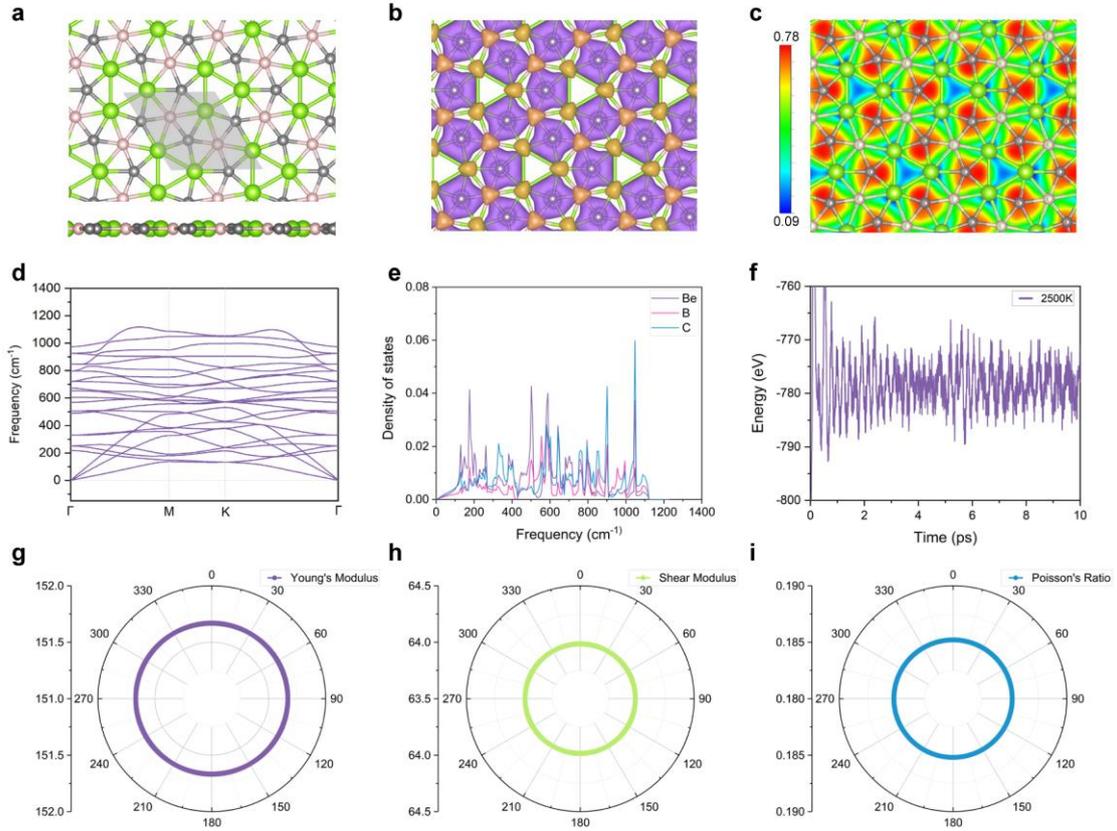

**Figure. 2. Stability analyses for the Be$_3$B$_2$C$_3$ monolayer**. (**a**) Vertical and lateral views of the geometric structure of Be$_3$B$_2$C$_3$ monolayer. The primitive unit cell is marked by grey diamond. Green, pink and grey spheres represent beryllium, boron and carbon atoms, respectively. (**b**) Deformation charge density of Be$_3$B$_2$C$_3$ monolayer. The isosurface value is 0.01 e Bohr$^{-3}$, where the electron-rich and electron-deficient regions are colored in purple and orange respectively. (**c**) Electron localization function (ELF) map of Be$_3$B$_2$C$_3$ monolayer. (**d**)-(**e**) Phonon spectrum and the corresponding phonon density of states of Be$_3$B$_2$C$_3$ monolayer. (**f**) Ab initio molecular dynamic simulations of the energy evolution of 4 × 4 × 1 supercell Be$_3$B$_2$C$_3$ monolayer with time running for 10 ps under 2500 K. (**g**) The direction-dependent Young's modulus, (**h**) shear modulus, and (**i**) Poisson's ratio of Be$_3$B$_2$C$_3$ monolayer. The direction along $\theta = 0°$ corresponds to the *x* axis.

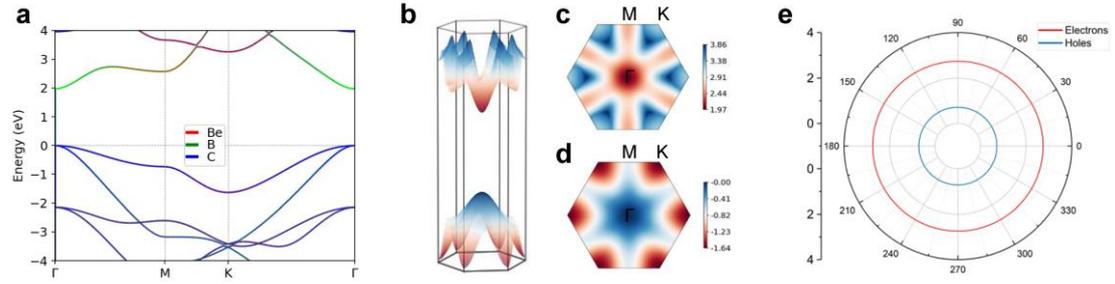

**Figure. 3. Electronical properties of Be$_3$B$_2$C$_3$ monolayer.** (**a**) The band structure of Be$_3$B$_2$C$_3$ monolayer by element weights calculated with HSE functional. (**b**) The 3D band edge structure of Be$_3$B$_2$C$_3$ monolayer in the first Brillouin zone. (**c**)-(**d**) The 2D band structure projection at CBM and VBM of Be$_3$B$_2$C$_3$ monolayer, respectively. (**e**) The direction-dependent Fermi velocity (×10$^5$ m/s) of electrons and holes for Be$_3$B$_2$C$_3$. $\theta = 0°$ corresponds to the $x$ axis.

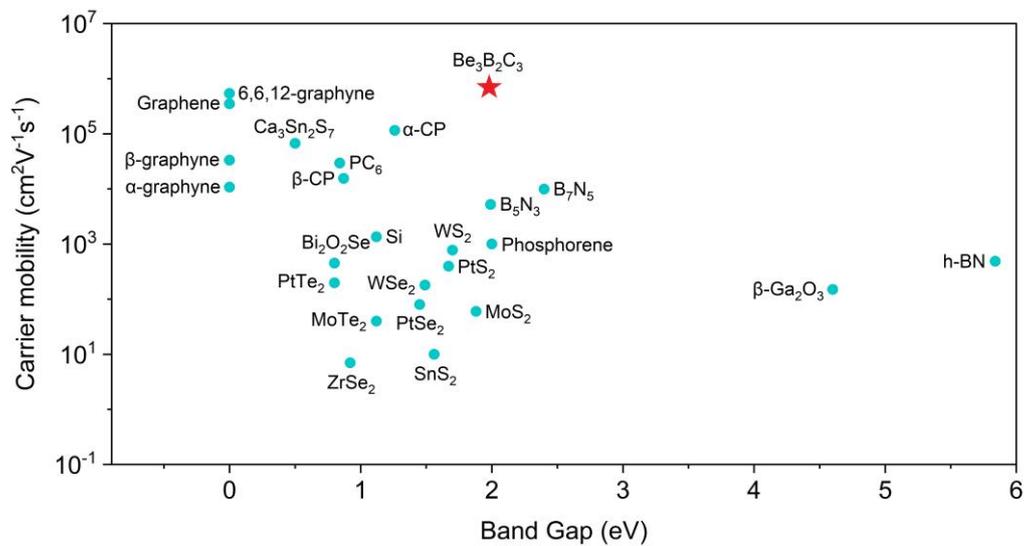

**Figure. 4. Summary of carrier mobility versus bandgap of the popular materials.**